\documentclass[preprint, 5p, twocolumn]{elsarticle}

\usepackage{amsmath,amssymb,bm,multirow,float,bbold}
\usepackage[american]{babel}
\allowdisplaybreaks

\newcommand{\lsim}{\stackrel{<}{_\sim}}

\newcommand{\be}{\begin{equation}}
\newcommand{\ee}{\end{equation}}

\def\beqn{\begin{eqnarray}}
\def\ba{\begin{array}{c}}
\def\bat{\begin{array}{cc}}
\def\bat{\begin{array}{cc}}
\def\ea{\end{array}}
\def\bat{\begin{array}{cc}}
\def\batt{\begin{array}{ccc}}
\def\eeqn{\end{eqnarray}}

\begin{document}
\begin{frontmatter}

\title{Effective Aligned 2HDM with a DFSZ-like invisible axion}

\author[a]{Alejandro Celis}
\ead{alejandro.celis@ific.uv.es}

\author[a]{Javier Fuentes-Mart\'{\i}n}
\ead{javier.fuentes@ific.uv.es}

\author[a]{Hugo Ser\^odio}
\ead{hugo.serodio@ific.uv.es}

\address[a]{Departament de F\'{\i}sica Te\`{o}rica and IFIC, Universitat de
Val\`{e}ncia-CSIC, E-46100, Burjassot, Spain}

\begin{abstract}
We discuss the possibility of having a non-minimal scalar sector at the weak scale within the framework of invisible axion models.  To frame our discussion we consider an extension of the Dine--Fischler--Srednicki--Zhitnitsky invisible axion model with two additional Higgs doublets blind under the Peccei-Quinn symmetry.    Due to mixing effects among the scalar fields, it is possible to obtain a rich scalar sector at the weak scale in certain decoupling limits of the theory.     In particular, this framework provides an ultraviolet completion of the so-called aligned two-Higgs-doublet model and solves the strong CP problem. The axion properties and the smallness of active neutrino masses are also discussed.

\end{abstract}

\end{frontmatter}

\section{Introduction \label{s:intro}}
The strong CP problem remains a puzzle of the $\mathrm{SU(3)}_{C} \otimes \mathrm{SU(2)}_{L} \otimes \mathrm{U(1)}_{Y}$ Standard Model (SM) gauge theory~\cite{Cheng:1987gp}.   The resolution of the $\mathrm{U(1)}_{A}$ problem by 't~Hooft noticing that the QCD vacuum is non-trivial~\cite{'tHooft:1976up}, led to the conclusion that the SM contains an additional source of CP violation besides the phase of the Cabibbo-Kobayashi--Maskawa (CKM) matrix~\cite{Cabibbo:1963yz}.  This additional source of CP violation is strongly constrained by present bounds on the neutron electric dipole moment~\cite{Baker:2006ts,Baluni:1978rf,Pospelov:2005pr}. 
 
One of the most compelling solutions to the strong CP problem involves the addition of a spontaneously broken Peccei-Quinn (PQ) $\mathrm{U(1)}_{\mbox{\scriptsize{PQ}}}$ symmetry to the theory~\cite{Peccei:1977hh}.  In this way one provides a dynamical solution to the strong CP problem while at the same time predicts the existence of a very light and long-lived pseudoscalar boson known as the axion~\cite{Weinberg:1977ma}.     The original PQ model with two-Higgs-doublets in which the PQ symmetry is broken at the electroweak (EW) scale was excluded long ago experimentally~\cite{Cheng:1987gp}.  By decoupling the breaking of the PQ symmetry from the breaking of the EW gauge symmetry one can build models with an \textit{invisible} axion which are still viable~\cite{Kim:1979if,Zhitnitsky:1980tq,Dias:2014osa}.    Invisible axion models avoid current experimental limits since the axion mass and couplings are suppressed by the PQ symmetry breaking scale, assumed to be much higher than the EW scale.  In the Dine--Fischler--Srednicki--Zhitnitsky  (DFSZ) invisible axion model, for example, one adds to the original PQ framework a complex scalar gauge singlet which acquires a large vacuum expectation value (vev)~\cite{Zhitnitsky:1980tq}. Besides solving the strong CP problem, invisible axion models can provide also a cure to other problems of the SM.  The invisible axion, for example, can be regarded as a well motivated cold dark matter candidate~\cite{Preskill:1982cy}.  

The recent discovery of a SM-like Higgs boson with mass around 125 GeV by the ATLAS~\cite{Aad:2012tfa} and CMS~\cite{Chatrchyan:2012ufa} collaborations represents an enormous achievement in particle physics and stands as a remarkable confirmation of the SM.  No fundamental principle of the SM forbids the presence of additional scalar fields related to the spontaneous breaking of the EW gauge symmetry.  Direct searches for additional scalars will then constitute an important part of the experimental program at the Large Hadron Collider during the following years.    

One of the simplest extensions of the SM scalar sector is the two-Higgs-doublet model (2HDM), which can lead to a very rich phenomenology~\cite{Branco:2011iw}.  However, such simple extension gives rise to unwanted flavor-changing neutral current (FCNC) interactions, which have to be suppressed in order to avoid conflict with experimental data.    One possible way out is to assume that the model is in a decoupling regime, in this case a SM-like Higgs remains at the weak scale while all the scalars with dangerous couplings become very heavy.   This is for example what usually happens in invisible axion models, where the large PQ scale brings the scalar sector to a decoupling limit.   A more interesting scenario from the phenomenological point of view is that an underlying symmetry is forbidding the dangerous FCNCs, leaving open the possibility of additional scalar fields at the weak scale.  This can be realized within the context of the 2HDM through the introduction of a discrete symmetry, leading to natural flavor conservation (NFC)~\cite{Glashow:1976nt}. Another possibility is requiring the alignment in flavor space of the Yukawa matrices~\cite{Pich:2009sp}. The so-called aligned two-Higgs-doublet model (A2HDM) contains as particular cases the different versions of 2HDMs with $\mathcal{Z}_2$ symmetries while at the same time introduces new sources of CP violation beyond the CKM phase. The DFSZ invisible axion model is actually built over a 2HDM for which NFC is imposed by the PQ symmetry. In the A2HDM, the Yukawa alignment condition is not imposed by any symmetry and is therefore spoiled by quantum corrections~\cite{Pich:2009sp,Ferreira:2010xe}.   Embedding the scalar sector of the A2HDM within a renormalizable invisible axion model is therefore not obvious and has not been achieved previously. 

In this paper we discuss the possibility of having invisible axion models with a non-minimal scalar sector at the EW scale. To frame our discussion, we consider in Sec.~\ref{s:frame} a simple extension of the DFSZ model with two additional Higgs doublets that are blind to the PQ symmetry.   The properties of the invisible axion are discussed in Sec.~\ref{sec:ax}. A study of the possible decoupling limits of this model is given in Sec.~\ref{s:mixing}.   Due to mixing effects among the scalar fields the decoupling structure of the theory becomes more rich than in the DFSZ model. We will show that in certain cases it is even possible to arrive to an effective 2HDM with a Yukawa aligned structure.  While the number of fields that are blind to the PQ symmetry could be reduced to just one for many of the issues discussed, by having two of these fields we guarantee that the scalar potential of the effective theory at the weak scale will be the most general one.  In Sec.~\ref{sec:rN} we present two ways of implementing small neutrino masses. We conclude in Sec.~\ref{sec:con}.

\section{Framework \label{s:frame} }  
We consider the DFSZ invisible axion model with two additional complex Higgs doublets that are not charged under the PQ symmetry.    The scalar sector of the model contains then four complex Higgs $\mathrm{SU(2)}_L$ doublets with hypercharge $Y= 1/2$ and a complex scalar gauge singlet $S$.  We denote by $\Phi_{1,2}$ the Higgs doublets that carry a PQ charge, the doublets that are blind to the PQ symmetry will be denoted by $\phi_{1,2}$.   All the Higgs doublets take part in the spontaneous breaking of the EW gauge symmetry by acquiring vevs $\langle \Phi_j^0 \rangle = u_j/\sqrt{2}$ and $\langle \phi_j^0 \rangle = v_j/\sqrt{2}\,$ ($j =1, 2$), with $ ( u_1^2 + \cdots + v_2^2  )^{1/2} \equiv v = (\sqrt{2}  G_F)^{-1/2}$ being fixed by the massive gauge boson masses.  As in the DFSZ model we assume that the global $\mathrm{U(1)}_{\mbox{\scriptsize{PQ}}}$ symmetry is spontaneously broken by a very large vev of the scalar field $\langle S \rangle = v_{\mbox{\scriptsize{PQ}}}/\sqrt{2}$ ($ v_{\mbox{\scriptsize{PQ}}}   \gg v$).  

Our scalar content will transform under the PQ symmetry as
\be
S\rightarrow e^{i  X_S  \,    \theta} S\,,\quad \Phi_j\rightarrow e^{iX_j\, \theta} \Phi_j\,,\quad \phi_j\rightarrow \phi_j \,.
\ee
The most relevant terms in the scalar potential, as will be explained in Sec.~\ref{s:mixing}, are the trilinear interactions
\be\label{eq:triplec}
\mu_{1,j}\Phi_1^\dagger \phi_j S\quad \text{and}\quad \mu_{2,j}\Phi_2^\dagger \phi_j S^\ast\,,
\ee
the implicit sum on $j=1,2$ is assumed. The parameters $\mu_{1(2),j}$ have mass dimension and determine the size of the mixing between both types of doublets. The above interactions lead to the following charge constraints
\be\label{eq:X1X2}
 X_1=-X_2= X_S\,.
\ee
The full scalar potential, built with the above constraints, is presented in the appendix. The PQ charge normalization is unphysical and therefore we shall set $X_S=1$, as it is usually done. 

In the Yukawa interactions we shall only couple the doublets $\Phi_j$, we call them active fields. The doublets $\phi_j$ will not couple to the fermions and thus we call them passive fields.   For simplicity, we choose the left-handed quark doublets to be blind under the chiral $\mathrm{U(1)}_{\mbox{\scriptsize PQ}}$. The charge assignments for the fermions are 
\begin{align}\nonumber
&Q_{L\alpha}\rightarrow Q_{L\alpha}, & \ell_{L\alpha}\rightarrow e^{iX_\ell\, \theta} \ell_{L\alpha},\\
&u_{R\alpha}\rightarrow e^{iX_u\, \theta} u_{R\alpha}, & e_{R\alpha}\rightarrow e^{iX_e\, \theta} e_{R\alpha} \,,\\\nonumber
&d_{R\alpha}\rightarrow e^{iX_d\, \theta} d_{R\alpha} \,.
\end{align}
Here $\alpha = \{1, 2, 3\}$ is a family index. The Yukawa Lagrangian reads
\be  \label{eq:lagY}
-\mathcal{L}_{Y}=\overline{Q_{L}}\,\Gamma \,\Phi_1 d_{R}+\overline{Q_{L}}\,\Delta \,\tilde{\Phi}_2 u_{R}+\overline{\ell_{L}}\,\Pi \,\Phi_k e_{R}+\text{h.c.}\,,
\ee
where $\tilde \Phi_2 \equiv i \sigma_2 \Phi_2^\ast$ with $\sigma_2$ the Pauli matrix.   The mass matrices for the fermions in the flavor basis are given by
\be
M_{d} = \frac{   \Gamma \, u_1 }{\sqrt{2}}  \,, \qquad M_{u} = \frac{   \Delta\, u_2 }{\sqrt{2}}    \,, \qquad M_{e} = \frac{   \Pi \, u_k }{\sqrt{2}}  \,.
\ee
The Yukawa interactions in Eq.~\eqref{eq:lagY} impose the charge constraints
\be
X_d=-X_1\,,\quad X_u=X_2\,,\quad X_e= X_\ell -X_k \,.
\ee
The charge $X_\ell$ is seen as a free parameter. Depending on the values of $k$, we will have different implementations of the natural flavor conservation (NFC) condition~\cite{Glashow:1976nt,Branco:2011iw}: $k=1$ (Type-II); and $k=2$ (Flipped).

These are just the usual implementations of NFC in the DFSZ model. Other implementations of the NFC condition, i.e. Type-I and Lepton-specific~\cite{Branco:2011iw}, where both up and down sectors couple to the same scalar doublet, would not solve the strong CP problem and are not considered.

\section{Axion properties} \label{sec:ax}
The low-energy effective interaction Lagrangian for the axion ($a$) can be written as\footnote{Here the substitution $f_a \simeq v_{\mbox{\tiny PQ}}$, with $f_a$ the axion decay constant and where terms of order $\mathcal{O}(v^2/v_{\mbox{\tiny PQ}})$ are neglected, is understood.}
\begin{align}
\mathcal{L}_a \supset&   \frac{\alpha}{8\pi v_{\mbox{\scriptsize PQ}}} C_{ag}C_{a\gamma}^{\mbox{\footnotesize eff}} \,a \, F_{\mu\nu}  \widetilde{F}^{\mu\nu}  +\frac{1}{2}C_{ae}  \frac{\partial_\mu a}{v_{\mbox{\scriptsize PQ}}} \, \overline{e}\gamma^\mu\gamma_5e\,.
\end{align}
The axion coupling to photons takes the form \cite{Srednicki:1985xd}
\be  \label{caf}
C_{a\gamma}^{\mbox{\footnotesize eff}} \simeq \frac{C_{a\gamma}}{C_{ag}}-\frac{2}{3}\frac{4+z}{1+z},
\ee
where the second term in $C_{a\gamma}^{\mbox{\footnotesize eff}}$ is a model-independent quantity which comes from the mixing of the axion with $\pi^0$ and $\eta$. The quantity $z$ is the quark mass ratio $z=m_u/m_d \simeq 0.56$, while $C_{ag}$ and $C_{a\gamma}$ are model-dependent quantities associated to the axial anomaly. In our model they read 
\begin{align}
\begin{split}
C_{ag}=&\left(X_1-X_2\right)N_f=6,\\
C_{a\gamma}=&2N_f\left(\frac{X_1}{3}-\frac{4X_2}{3}+X_k\right)=\left\{
\begin{array}{ll}
16\,, & k=1\\
\;\,4\,, & k=2
\end{array}\right. \,.
\end{split}
\end{align}
\noindent Here $N_f =3$ is the number of fermion families and $X_i$ represent the PQ charges of the active scalar doublets. The axion mass is given by~\cite{Weinberg:1977ma}
\be
m_a \simeq \frac{f_\pi m_\pi C_{ag}}{v_{\mbox{\scriptsize PQ}}} \frac{\sqrt{z}}{1+z }   \simeq 36~\text{meV} \times \left( \frac{  10^9~\text{GeV}  }{  v_{\mbox{\scriptsize PQ}} }  \right) \,,
\ee
with $m_\pi \simeq 135$~MeV and $f_\pi \simeq 92$~MeV the pion mass and decay constant, respectively. 

\begin{figure}[t]
\includegraphics[width=8cm]{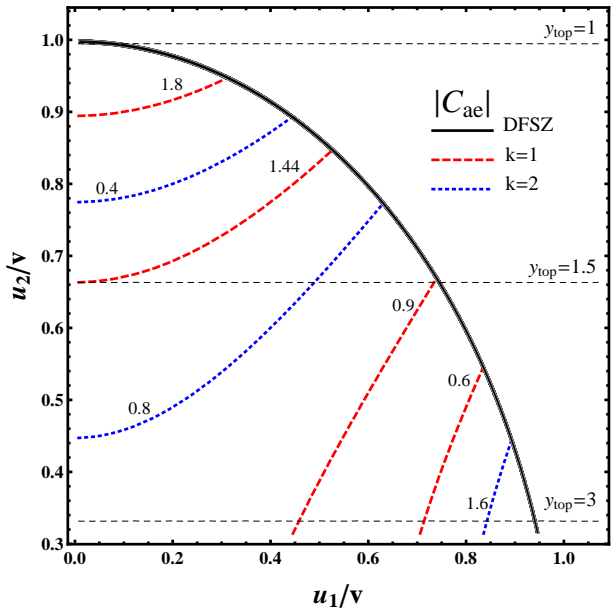}
\caption{Axion coupling to electrons, $|C_{ae}|$, in the $u_2/v$ vs. $u_1/v$ plane. The DFSZ model is represented by the solid black line, our framework with red dashed line for $k=1$ and blue dotted one for $k=2$.   Very small values of $u_2/v$ would lead to a non-perturbative top Yukawa and are not shown.  }
\label{fig:Cae}
\end{figure}

So far, the axion properties shown are exactly the same than in the DFSZ model. However, in the computation of the axion couplings to matter one has to modify the axion current in order to take into account the spontaneous breaking of the EW gauge symmetry. That is, one has to define the axion so that it does not mix with the Goldstone boson of the $Z$. Since the information on the passive fields enter through the neutral Goldstone boson our model will differ on the axion couplings to matter. As a result, the PQ-charges are modified in the following way~\cite{Srednicki:1985xd}
\begin{align}  \label{eq.orto}
X_{k}'&=X_{k}-\frac{1}{v^2}\sum_{m=1}^{2}  u_m^2   X_m\,,
\end{align}
with $k=1,2$. Therefore, the axion coupling to electrons is given by
\begin{align}\renewcommand\arraystretch{2}  \label{cae}
C_{ae}=X_k'=\left\{\begin{matrix}
               2\dfrac{u_2^2}{v^2}+\dfrac{v_1^2+v_2^2}{v^2} & \mbox{for }k=1\\
               -2\dfrac{u_1^2}{v^2}-\dfrac{v_1^2+v_2^2}{v^2} & \mbox{for }k=2
              \end{matrix}\right. \,,
\end{align}
\noindent where Eq. (\ref{eq:X1X2}) has been used.   As expected, we recover the DFSZ result for the axion properties in the limit $v_{1,2} = 0$, when the passive fields do not take part in the breaking of the EW symmetry. However, we can have significant deviations when this is not the case. In Fig.~\ref{fig:Cae} we plot the absolute value of the axion-electron coupling, $|C_{ae}|$, in terms of the ratios $u_k/v$. The black solid line corresponds to the DFSZ scenario, while the dashed red and dotted blue lines to our cases $k=1$ and $k=2$, respectively. Let us take the $k=1$ implementation as an example; the dashed red lines are contours and their intersection with the solid black line give the $|C_{ae}|$ value in the DFSZ model. Fixing, for example, the top Yukawa to $y_{top}=1.5$ (horizontal dashed line) the DFSZ scenario gives $|C_{ae}|=0.9$. However, this horizontal line crosses not only one dashed red contour but many. In particular, for this specific $y_{top}$ we have $|C_{ae}|\in [0.9\,,\,1.44]$. This allow us to increase the axion-electron coupling up to $60\%$. For the scenario $k=2$ the inverse happens, i.e. we can decrease the axion-electron coupling. The adimensional axion-electron coupling constant is defined as
\be
|g_{ae}|=\frac{m_e |C_{ae}|}{v_{\mbox{\scriptsize PQ}}}\simeq 1.4\times 10^{-14}\times \left(\frac{m_a}{\text{meV}}\right)\times |C_{ae}|\,.
\ee
The axion emission from white dwarfs and stellar evolution considerations introduce the strongest bound on axion-electron interactions, requiring $|g_{ae}|\lsim 1.3\times 10^{-13}$~\cite{Raffelt:2006cw,Bertolami:2014wua}. This leads to the mass bound
\be
m_a|C_{ae}|\lsim\,10 \,\text{meV}\,.
\ee
Taking the scenario $k=1$ and fixing the value of the top Yukawa, we see that the axion mass is more constrained as we depart from the DFSZ limit. For the scenario $k=2$ the inverse happens, as we depart from the DFSZ limit we soften the bound on the axion mass (for a fixed value of the top Yukawa). Therefore, the presence of the passive fields can have important implications for the energy-loss in stars by modifying the axion coupling to electrons.

Taking into account perturbativity of the top Yukawa, the allowed range for $|C_{ae}|$ is roughly $[0.2, 2]$ and $[0,1.8]$ for $k=1$ and $k=2$, respectively. This implies the following bound on the axion mass $m_a\lsim 5\,\text{meV}$.   Such bound is well compatible with the region where the invisible axion could constitute all of the dark matter in the Universe, see Ref.~\cite{Dias:2014osa} and references therein.

The presence of the passive fields would also give rise to similar modifications of the axion coupling to hadrons~\cite{Srednicki:1985xd}, relevant for interpreting the supernova SN 1987A limits~\cite{Raffelt:2006cw}.     The passive fields, on the other hand, do not change the axion coupling to photons. In our scenario this implies that bounds relying on the axion coupling to photons would be the same than in the DSFZ model.  In particular, constraints from the Solar age, helioseismology, the Solar neutrino flux as well as direct axion searches via axion-photon conversion are not sensitive to the passive fields~\cite{Raffelt:2006cw}.

\section{Mixing active/passive doublets and the decoupling limit\label{s:mixing}}
A distinctive feature of invisible axion models is that the large PQ symmetry breaking scale usually brings the scalar sector to a decoupling scenario.     A SM-like Higgs remains at the weak scale while the other scalar fields (with the exception of the axion) get masses around $v_{\mbox{\scriptsize{PQ}}}$.   In the DFSZ model for example, decoupling arises due to terms in the scalar potential mixing the Higgs doublets with the scalar singlet; these terms are crucial so that the axion actually becomes invisible.  Under specific circumstances one can avoid decoupling in the DFSZ model and have the two Higgs doublets at the weak scale, protection against dangerous flavor changing scalar couplings is guaranteed by the NFC condition.  In order to illustrate the decoupling limit, let us consider the DFSZ scalar potential which is a particular case of our more general scalar sector, where the passive fields are absent, i.e. $V_{\mbox{\footnotesize{DFSZ}}}\;=\;\left.V\right|_{\phi_j=0}$.
Due to the large hierarchy on the vevs, i.e. $ v_{\mbox{\scriptsize{PQ}}}   \gg v$, we can extract 
\be
v_{\mbox{\scriptsize{PQ}}}^2 = - 2\mu_S^2/ \lambda_S + \mathcal{O}(v^2)\,.
\ee
Up to $\mathcal{O}(v^2)$, we can deal with this mixing as being $\mathrm{SU(2)}_L$ preserving.   The mass matrix for the doublets reads
\begin{align}
V^{\mbox{\footnotesize mass}}_{\mbox{\footnotesize{DFSZ}}} \;=\;   \Phi_i ^{\dag}  \, \left(\mathcal{  M}_A\right)_{ij} \, \Phi_j  + \mathcal{O}(v^2)       \,,  
\end{align}
with $\mathcal{  M}_A$ given in Eq.~\eqref{eq:MA}. The decoupling condition can be readily obtained by going to the Higgs basis in which only one Higgs doublet takes a vev,  the Higgs doublet that does not acquire a vev will decouple if 
\be
\frac{| \lambda_{12}^{\Phi S}|  \, v_{\mbox{\scriptsize{PQ}}}^2}{  2 \cos \beta \sin \beta }  \gg v^2  \,.
\ee
Here $\tan \beta \equiv  \langle \Phi_2^0 \rangle/\langle \Phi_1^0 \rangle$ and $\lambda_{12}^{\Phi S}$ is defined in Eq.~\eqref{eq:ScalarPot}.  In the decoupling limit the Higgs doublet that gets a vev remains at the EW scale: three degrees of freedom of this doublet correspond to the Goldstone bosons that give mass to the massive gauge vector bosons while the remaining degree of freedom is a SM-like Higgs boson.       If $|\lambda_{12}^{\Phi S}|$ happens to be small enough, both doublets remain at the weak scale and a plethora of new physics phenomena associated with the Higgs sector becomes accessible to experiments.  

In the framework presented in Sec.~\ref{s:frame}, the breaking of $\mathrm{U(1)}_{\mbox{\scriptsize{PQ}}}$ by the large vev $v_{\mbox{\scriptsize{PQ}}}$ induces a non-negligible mixing between the active and passive scalar doublets.  The decoupling structure of this framework will then be richer than in the DFSZ model.    Defining the scalar field $\varphi=(\Phi_1,\, \Phi_2,\, \phi_1,\, \phi_2)^T$, we want to diagonalize the mass terms for the doublets $\varphi_i^{\dag}  \mathcal{M}_{ij}  \, \varphi_{j}$, where
\be  \label{eqmas}
\mathcal{M} \;=\;   \left( \begin{array}{cc }    \mathcal{M}_{A} &  \mathcal{M}_{B}  \\   \mathcal{M}_{B}^{\dag}    & \mathcal{M}_{C}  \end{array}  \right) \,.
\ee
Here $\mathcal{M}$ is a $4 \times 4$ hermitian matrix and the specific form of the $2 \times 2$ blocks $\mathcal{M}_{A,B,C}$ is given in the appendix. The block $\mathcal{M}_B$ is responsible for the mixing of active and passive fields, it comes solely from the interaction in Eq.~\eqref{eq:triplec}. Let us denote by $H_j$ ($j= 1, \ldots, 4$) the mass eigenstates, ordered from the heaviest to the lightest one ($|M_{H_m}| \geq |M_{H_n}|$ for $m < n$). We must find the unitary transformation $\mathcal{R}$, i.e. $\varphi_i = \mathcal{R}_{ij} \,  H_j$, that makes $\mathcal{M}$ diagonal. The Yukawa interactions will contain, in general, the four mass eigenstates $H_j$ coupling to the fermions. 

We are interested in the two following decoupling limits: (1) $H_{3,4}$ at the weak scale; (2) $H_{4}$ at the weak scale. In case (2) we will get a SM-like Higgs sector at the weak scale. Therefore, we shall focus on case $(1)$ (case (2) can be seen as a limiting case). Working in the decoupling limit $(1)$, we do not need the full information on the entries of $\mathcal{R}$ in order to check the low energy Yukawa interactions. The relevant entries are the block that mixes the active fields with the lightest mass eigenstates, that is
\be
\begin{pmatrix}     \label{relationJ}
\Phi_1\\
\Phi_2
\end{pmatrix}
=\widehat{\mathcal{R}} 
\begin{pmatrix}
H_3\\
H_4
\end{pmatrix}\,,\quad\text{with}\quad
\widehat{\mathcal{R}} =
\begin{pmatrix}
\mathcal{R}_{13}&\mathcal{R}_{14}\\
\mathcal{R}_{23}&\mathcal{R}_{24}
\end{pmatrix}\,.
\ee
The matrix $\widehat{\mathcal{R}}$ is in general not unitary.   The effective Yukawa interactions will be given by
\begin{align}
\begin{split}
-\mathcal{L}^{\mbox{\footnotesize{eff}}}_{Y}\;=\;&\overline{Q_{L}}\,  \Gamma \,(\mathcal{R}_{13}H_3 +\mathcal{R}_{14}H_4)d_{R}\\
&+\overline{Q_{L}}\,\Delta \,(\mathcal{R}_{23}^\ast\tilde{H}_3 +\mathcal{R}_{24}^\ast \tilde{H}_4) u_{R}\\
&+\overline{\ell_{L}}\, \Pi \,(\mathcal{R}_{k3}H_3 +\mathcal{R}_{k4}H_4) e_{R}+\text{h.c.}\,
\end{split}
\end{align}
In the decoupling limit, the EW vev should reside completely in the light doublets $\langle H^0_{3,4} \rangle =w_{3,4}/\sqrt{2}$, with $(w_3^2+w_4^2)^{1/2} = v$.     We describe below the Yukawa structures that can arise at the weak scale for different forms of the matrix $\widehat{\mathcal{R}}$.     The entries denoted by $\times$ shall represent not only nonzero entries, but also of $\mathcal{O}(1)$. The last requirement guarantees perturbative Yukawa couplings. We then have the following cases:
\begin{itemize}

\item $\widehat{\mathcal{R}}=
\begin{pmatrix}
\times& 0 \\
\times &\times
\end{pmatrix}
$

A mixing matrix $\widehat{\mathcal{R}}$ with this structure will give rise to Yukawa alignment~\cite{Pich:2009sp} in the effective theory.   A possible texture for the mass matrix is
\begin{align}
\begin{split}
&\frac{\mathcal{M}_{A}}{2b}\sim
\begin{pmatrix}
1+\dfrac{\epsilon}{2b}&1\\
1&1
\end{pmatrix}\,,\quad \frac{\mathcal{M}_{B}}{b}\sim
\begin{pmatrix}
1&1\\
1&1
\end{pmatrix}\,,\\
 &\frac{\mathcal{M}_{C}}{b}\sim
\begin{pmatrix}
2&-1\\
-1&2
\end{pmatrix}\,,
\end{split}
\end{align}
with $b$ and $\epsilon$ being parameters of $\mathcal{O}(b)\sim\mathcal{O}(v_{\mbox{\scriptsize{PQ}}}^2)\gg \mathcal{O}(\epsilon)$.    The mass spectrum is of the form, up to $\mathcal{O}(v^2)$,
\be
M_{H_{1}}^2\sim 5b\,,\, M_{H_{2}}^2\sim 3b\,,\,M_{H_3}^2\sim \epsilon\,,\, M_{H_4}^2\sim 0\,.
\ee
Since one is free to perform a basis transformation among the light Higgs doublets, other forms of the mixing matrix $\widehat{\mathcal{R}}$ leading to Yukawa alignment at the weak scale are equivalent to the one presented previously.     In this framework one can only obtain two independent alignment parameters, contrary to the most general hypothesis of Yukawa alignment formulated in Ref.~\cite{Pich:2009sp} with three independent alignment parameters.

\item $\widehat{\mathcal{R}}=
\begin{pmatrix}
\times & 0 \\
\times & 0 
\end{pmatrix}$

Any UV implementation will always lead in this case to an effective Type-I scenario, where all the fermions couple to the same doublet at the weak scale. A possible texture for the mass matrix is
\be
\frac{\mathcal{M}_A}{b}\sim
\begin{pmatrix}
1+\dfrac{\epsilon}{b}&1\\
1&1+\dfrac{\epsilon}{b}
\end{pmatrix},\; \frac{\mathcal{M}_{B}}{c}\sim\frac{\mathcal{M}_{C}}{b}\sim
\begin{pmatrix}
1&1\\
1&1
\end{pmatrix}.
\ee
Here $c$ is a parameter of $\mathcal{O}(c)\sim\mathcal{O}(b)\sim\mathcal{O}(v_{\mbox{\scriptsize{PQ}}}^2)\gg \mathcal{O}(\epsilon)$.   This leads to a spectrum of the form
\be
M_{H_{1,2}}^2\simeq 2(b\pm c)\,,\, M_3^2\simeq \epsilon\,,\, M_4^2\simeq 0\,,
\ee
with the same hierarchy as before.
\end{itemize}

In the previous cases, the original Yukawa structure of the active fields is not manifest at the weak scale.   A large mixing $\mu_{1(2),j} \sim v_{\mbox{\scriptsize{PQ}}}$ between the active and passive fields generates a decoupling scenario where the light scalar states contain a significant admixture of both types of fields.      However, a large mixing between active and passive fields does not guarantee that the Yukawa structure will be different in the effective theory.

\begin{itemize}
\item $\widehat{\mathcal{R}}=
\begin{pmatrix}
\times&0\\
0&\times
\end{pmatrix}$

The original UV implementation will remain at the effective level.   A possible texture for the mass matrix is
\begin{align}
\begin{split}
&\frac{\mathcal{M}_{A}}{b}\sim
\begin{pmatrix}
1+\dfrac{\epsilon}{b}&0\\
0&1
\end{pmatrix}\,,\quad \frac{\mathcal{M}_{B}}{b}\sim
\begin{pmatrix}
0&1\\
1&0
\end{pmatrix}\,,\\
 &\frac{\mathcal{M}_{C}}{b}\sim
\begin{pmatrix}
1&0\\
0&1
\end{pmatrix}\,.
\end{split}
\end{align}
We get the mass spectrum
\be
M_{H_{1, 2}}^2\sim 2b\,,\, M_{H_3}^2\sim \epsilon\,,\, M_{H_4}^2\sim  0  \,.
\ee
In general, the original Yukawa structure in Eq.~\eqref{eq:lagY} will remain in the effective theory if the mixing matrix $\widehat{\mathcal{R}}$ can be brought into diagonal form by a basis transformation of the light doublets.  Finally, if the mixing between the active and passive doublets is negligible $\mu_{1(2),j} \ll v_{\mbox{\scriptsize{PQ}}}$, the only way to get the desired decoupling is that the light fields $H_{3,4}$ are simply two independent combinations of the active doublets.  In this case the Yukawa structure is not altered.       
 \end{itemize}

As noted in Sec.~\ref{s:frame}, only the Type-II or Flipped implementations can solve the strong CP problem.   However, in our scenario we are able to mimic the DFSZ axion and still allow (at the effective level) for a Type-I, Type-II, flipped or even aligned Yukawa structure. Also, recall that in contrast with the usual two-Higgs-doublet models with NFC, our effective scalar potential has the most general form. Finally, we are not able to reproduce (at the effective level) the fepton-specific scenario since the active field coupling to the charged leptons at the UV level is always one of the active fields coupling to the up or down quarks.

\section{Adding right-handed neutrinos}  \label{sec:rN}
We shall work in the canonical extension with three right-handed neutrinos $N_{R\alpha}$ ($\alpha=1,2,3$). Since the Dirac or Majorana nature of the light neutrinos is still unknown we shall present an implementation for both scenarios.

The Majorana neutrinos can be implemented in the well known Type-I seesaw mechanism~\cite{Minkowski:1977sc}. In this framework the fermionic interaction Lagrangian gets extended by
\be\label{seesawI}
-\mathcal{L}_{\nu}=\overline{\ell_L} Y \, \tilde{\Phi}_r N_R+\overline{N_R^c} \,A \, N_R \,S+\text{h.c.}\,
\ee
Here $r=1\text{ or }2$ depending on the implementation of the NFC condition and $A$ is a dimensionless $3 \times 3$ symmetric complex matrix. We need to impose a non-trivial transformation of $N_R$ under the $\mathrm{U(1)}_{\mbox{\scriptsize{PQ}}}$ symmetry. With the field transformation
\be
N_{R\alpha}\rightarrow e^{iX_N\theta}N_{R\alpha}\,,
\ee
the above Lagrangian implies the charge constraints\footnote{Allowing instead for the term $\overline{N_R^c} \,A \, N_R \,S^*$ in Eq.~\eqref{seesawI} would imply $X_N = \frac{1}{2} $. }
\be
X_N=-\frac{1}{2}\,,\quad X_\ell= X_N -X_r\,.
\ee
After the breaking of the PQ-symmetry, a mass term for the right-handed fields is generated and the resulting low-energy neutrino mass matrix will then be given by
\be\label{mnutypeI}
m_\nu \simeq  - \frac{u_r^{\ast 2}}{  2\sqrt{2}  v_{\mbox{\scriptsize{PQ}}}} \,Y \, A^{-1} \, Y^T \,.
\ee
If in Eq.~\eqref{seesawI} instead of $\Phi_r$ we had the passive doublets coupling to neutrinos, $X_\ell$ becomes $-1/2$ and in Eq.~\eqref{mnutypeI} we must do the replacement $u_r^\ast Y\rightarrow \sum_j v_j^\ast Y_j$.

In general, the introduction of a Majorana mass term for the right-handed neutrinos breaks lepton number. The presence of a complex scalar singlet allows the definition of a conserved lepton number $\mathrm{U(1)}_L$ in Eq.~\eqref{seesawI}, where all leptons have associated a $+1$ charge ($-1$ for anti-leptons) and the complex scalar $S$ a $-2$ charge~\cite{Chikashige:1980ui}. However, the presence in the scalar potential of the interaction terms in Eq.~\eqref{eq:triplec} explicitly violates lepton number.  One could see these type of interactions as a soft breaking, allowing lepton number conservation in a natural limit. As explained in Ref.~\cite{Chikashige:1980ui}, in that symmetric limit we get a majoron (the Goldstone of the $\mathrm{U(1)}_L$). In these models the majoron can transmute into the invisible axion as the soft symmetry breaking term is turned on. Our scenario is a bit different, since we cannot define a lepton number up to a soft breaking.  The trilinear terms in Eq.~\eqref{eq:triplec} must be close to the PQ scale in order to obtain a Yukawa aligned structure at the weak scale while avoiding non-perturbative Yukawa couplings at the same time.   Besides, the interaction terms in Eq.~\eqref{seesawI}, we also have the dimension four lepton number violating term $\Phi_1^\dagger \Phi_2 S^2$.\footnote{This term can be eliminated choosing a different PQ charge assignment, in that case the trilinear couplings are promoted to a dimension four term.} In this way, lepton number is not at all softly broken. 

Summing up, in our scenario lepton number is explicitly (and not softly) broken and the invisible axion will have no remnant of a majoron. Therefore, the seesaw scale can be related with the PQ scale, but the dynamical origin of lepton number violation is not approached in this model.

Choosing the charge assignment $X_\ell=X_N\neq 0,\pm1/2$, we can avoid the Majorana mass term for the right-handed neutrinos as well as their Yukawa coupling with the active doublets.    In this case the neutrinos obtain Dirac masses from their Yukawa interaction with the passive fields, 
\be\label{dirac}
-\mathcal{L}_{\nu}=\overline{\ell_L} Y_j \, \tilde{\phi}_j N_R+\text{h.c.}\,
\ee
The neutrino mass matrix will be given by 
\be
m_\nu= \frac{v_j^\ast}{\sqrt{2}}Y_j\,.
\ee
This scenario is not as popular as the seesaw mechanism for two main reasons: the requirement of very small Yukawa couplings, without any dynamical origin; and the need for a new imposed symmetry, a global $B-L$.

In our framework the very small Yukawa couplings can be avoided if we are working near the DFSZ limit. In this limiting case we have the strong hierarchy $\mathcal{O}(u_j)\gg \mathcal{O}(v_j)$, allowing the neutrino Yukawas to be as tuned as the charged lepton ones. Concerning the global $B-L$ symmetry; due to the particular charge assignments under $\mathrm{U(1)}_{\mbox{\scriptsize{PQ}}}$, the theory posses an accidental $B-L$ global symmetry which remains unbroken. Note that, while the previous seesaw scenario can be implemented in the usual DFSZ model, the Dirac case is only possible (without resorting to very small Yukawa couplings) if the scalar fields coupling to neutrinos do not couple to other type of matter, i.e. are passive fields.

\section{Conclusions}  \label{sec:con}
In this work we have considered the DFSZ invisible axion model with an additional pair of Higgs doublets that are blind to the PQ symmetry.    Due to mixing effects among the scalar fields it is possible to obtain a rich scalar sector at the weak scale with an underlying natural flavor conservation condition which guarantees the absence of dangerous flavor changing scalar couplings.   We have shown that in a particular decoupling limit, two Higgs doublets remain at the weak scale with a Yukawa aligned structure~\cite{Pich:2009sp}, while all the other scalars (with the exemption of the axion) have masses close to the PQ symmetry breaking scale.   In this limit, the model can then be regarded as an ultraviolet (UV) completion of the so-called aligned two-Higgs-doublet model (A2HDM)~\cite{Pich:2009sp}.      Compared with the original formulation of the A2HDM, our framework posses some important differences. In our scenario a chiral global $\mathrm{U(1)}_{\mbox{\scriptsize{PQ}}}$ symmetry provides a dynamical solution to the strong CP problem via the PQ mechanism, with a DFSZ-like axion. On the other hand, our model contains at most two independent complex alignment parameters while the general A2HDM contains three (with the same fermionic content than the SM).  We can also extend it to accommodate small active neutrino masses with either Dirac or Majorana neutrinos.

Other models that can give rise to Yukawa alignment at the weak scale have been formulated~\cite{Bae:2010ai}, none of these however solve the strong CP problem.   Having an UV completion of the A2HDM that solves the strong CP problem is crucial for example when interpreting the stringent limits from hadronic electric dipole moments~\cite{Jung:2013hka}.

Finally, it is worth stressing that invisible axion models as the one proposed in this Letter have some drawbacks; besides the fact that the PQ symmetry and the particle content might seem ad-hoc. Being $\mathrm{U(1)}_{\mbox{\scriptsize{PQ}}}$ a continuous symmetry, gravitational effects can introduce large contributions to the axion mass and spoil the solution to the strong CP problem~\cite{Georgi:1981pu}.   Fortunately, there are loopholes which allow these kind of models to be protected against such effects. One possible solution is to have discrete symmetries of the type $\mathcal{Z}_N$ with large $N$ protecting the model.   These discrete symmetries could come for instance from a gauged $\mathrm{U(1)}$ symmetry, that is then broken down to a discrete subgroup at a very high energy, see Ref.~\cite{Dias:2014osa} and references therein.  The presence of these discrete symmetries can then induce in the scalar potential an accidental $\mathrm{U(1)}_{\mbox{\scriptsize{PQ}}}$ symmetry, making the PQ symmetry no longer seem ad hoc while protecting the solution to the strong CP problem~\cite{Georgi:1981pu}. While this is out of the scope of our work, we call the attention that in order to make this model more natural the previous mechanism, or an equivalent one, should be implemented.

\section*{Acknowledgements}
We thank Jorge Portol\'{e}s, Celso C. Nishi and Jordy de Vries for a careful reading of the manuscript and valuable comments.
The work of H.S. is funded by the European FEDER,
Spanish MINECO, under the grant FPA2011-23596, and the Portuguese FCT project
PTDC/FIS-NUC/0548/2012.  The work of A.C. and J.F. has been supported in part by the Spanish Government and ERDF funds from the European Commission [Grants FPA2011-23778 and CSD2007-00042 (Consolider Project CPAN)].   J. F. also acknowledges VLC-CAMPUS for an ``Atracci\'{o} del Talent"  scholarship.   

\begin{appendix}

\section*{Appendix: The Full Scalar Potential}
\label{app1} 
\renewcommand{\theequation}{A-\arabic{equation}}
The scalar potential for the Higgs doublets $\Phi_j, \phi_j $ ($j=1,2$) and the complex scalar gauge singlet $S$ can be written as $V = V_S + V_{\Phi} + V_{\phi}$, with
\begin{align}\label{eq:ScalarPot}
\nonumber V_S=&\mu_S^2|S|^2+\lambda_S|S|^4+\lambda_i^{\Phi S}(\Phi_i^\dagger \Phi_i)|S|^2\\
\nonumber &+\left[\lambda_{12}^{\Phi S}(\Phi_1^\dagger \Phi_2)S^2+\text{h.c.}\right]\\\
\nonumber &+\lambda_i^{\phi S}(\phi_i^\dagger \phi_i)|S|^2+\left[\lambda_{12}^{\phi S}(\phi_1^\dagger \phi_2)|S|^2+\text{h.c.}\right]\\
\nonumber &+\left[\mu_{1,i}\Phi_1^\dagger \phi_i S+\mu_{2,i}\Phi_2^\dagger \phi_i S^\ast+\text{h.c.}\right]\,,\\
\nonumber V_\Phi=&M_i^2 \Phi_i^\dagger\Phi_i+\lambda^\Phi_{ii,jj}(\Phi^\dagger_i\Phi_i)(\Phi_j^\dagger\Phi_j)\\
\nonumber &+\lambda_{ii,jj}^{\Phi\phi}(\Phi^\dagger_i\Phi_i)(\phi^\dagger_j\phi_j)+\lambda_{ii,jj}^{\prime\Phi\phi}(\Phi^\dagger_i\phi_j)(\phi^\dagger_j\Phi_i)\\
\nonumber &+\lambda^\Phi_{12,21}(\Phi^\dagger_1\Phi_2)(\Phi_2^\dagger\Phi_1)+\left[\lambda^{\Phi\phi}_{ii,12}(\Phi^\dagger_i\Phi_i)(\phi_1^\dagger\phi_2)\right.\\
\nonumber &\left.+\lambda^{\prime\Phi\phi}_{ii,12}(\Phi^\dagger_i\phi_2)(\phi_1^\dagger\Phi_i)+\text{h.c.}\right]\,,\\
V_\phi=&m_{ij}^2\phi^\dagger_i\phi_j+\frac{1}{2}\lambda_{ij,kl}(\phi^\dagger_i\phi_j)(\phi_k^\dagger\phi_l)\,,
\end{align}
with $\lambda_{ij,kl}=\lambda_{kl,ij}$, $m_{ij}^2=(m_{ji}^2)^\ast$ and $\lambda_{ij,kl}=\lambda_{ji,lk}^\ast$.  The blocks of the mass matrix for the Higgs doublets in Eq.~\eqref{eqmas} are then given by
\begin{subequations}\label{eq:M}
\begin{align}
\mathcal{M}_{A}&=  \left( \begin{array}{c c  }
    M_1^2 +  \,  \frac{\lambda^{\Phi S}_{1}}{2}v_{\mbox{\tiny{PQ}}}^2   &   \frac{\lambda^{\Phi S}_{12}}{2}v_{\mbox{\tiny{PQ}}}^2       \\[0.1cm]
\frac{(\lambda_{12}^{\Phi S})^\ast}{2}v_{\mbox{\tiny{PQ}}}^2       & M_2^2 +  \frac{\lambda^{\Phi S}_{2}}{2}v_{\mbox{\tiny{PQ}}}^2    \end{array}\right)\,,\label{eq:MA}\\
\mathcal{M}_{B}&=    \frac{1}{ \sqrt{2}  }v_{\mbox{\scriptsize{PQ}}}^2   \left( \begin{array}{c c  }
    \mu_{1,1}   &   \mu_{1,2}          \\
  \mu_{2,1}     &    \mu_{2,2}     \end{array}\right)\, ,\label{eq:MB}\\
\mathcal{M}_{C}&=  \left( \begin{array}{c c  }
    m_{11}^2 + \frac{\lambda_{1}^{\phi S}}{2}v_{\mbox{\tiny{PQ}}}^2      & m_{12}^2  +  \frac{ \lambda_{12}^{\phi S}}{2}v_{\mbox{\tiny{PQ}}}^2        \\[0.1cm]
 ( m_{12}^2)^* +\frac{(\lambda_{12}^{\phi S})^\ast}{2}v_{\mbox{\tiny{PQ}}}^2       & m_{22}^2 +\frac{ \lambda_{2}^{\phi S}}{2}v_{\mbox{\tiny{PQ}}}^2     \end{array}\right)  \,.  \label{eq:MC}
\end{align} 
\end{subequations}

\end{appendix}


\begin{thebibliography}{99}


\bibitem{Cheng:1987gp}
H.~-Y.~Cheng,
  Phys.\ Rept.\  {\bf 158} (1988) 1;
  R.~D.~Peccei,
  Lect.\ Notes Phys.\  {\bf 741} (2008) 3
  [hep-ph/0607268];
  J.~E.~Kim and G.~Carosi,
  Rev.\ Mod.\ Phys.\  {\bf 82} (2010) 557
  [arXiv:0807.3125 [hep-ph]].

\bibitem{'tHooft:1976up}
  G.~'t Hooft,
  Phys.\ Rev.\ Lett.\  {\bf 37} (1976) 8;
  G.~'t Hooft,
  Phys.\ Rev.\ D {\bf 14} (1976) 3432
   [Erratum-ibid.\ D {\bf 18} (1978) 2199].
  
\bibitem{Cabibbo:1963yz}
  N.~Cabibbo,
  Phys.\ Rev.\ Lett.\  {\bf 10} (1963) 531;
  M.~Kobayashi and T.~Maskawa,
  Prog.\ Theor.\ Phys.\  {\bf 49} (1973) 652.
    
\bibitem{Baker:2006ts}
  C.~A.~Baker, D.~D.~Doyle, P.~Geltenbort, K.~Green, M.~G.~D.~van der Grinten, P.~G.~Harris, P.~Iaydjiev and S.~N.~Ivanov {\it et al.},
  Phys.\ Rev.\ Lett.\  {\bf 97} (2006) 131801
  [hep-ex/0602020].
  
  
        
    
    
    
\bibitem{Baluni:1978rf}
  V.~Baluni,
  Phys.\ Rev.\ D {\bf 19} (1979) 2227;
  R.~J.~Crewther, P.~Di Vecchia, G.~Veneziano and E.~Witten,
  Phys.\ Lett.\ B {\bf 88} (1979) 123
   [Erratum-ibid.\ B {\bf 91} (1980) 487].


\bibitem{Pospelov:2005pr}
M.~Pospelov and A.~Ritz,
  Annals Phys.\  {\bf 318} (2005) 119
  [hep-ph/0504231].



\bibitem{Peccei:1977hh}
  R.~D.~Peccei and H.~R.~Quinn,
  Phys.\ Rev.\ Lett.\  {\bf 38} (1977) 1440;
  R.~D.~Peccei and H.~R.~Quinn,
  Phys.\ Rev.\ D {\bf 16} (1977) 1791.
  
  


\bibitem{Weinberg:1977ma}
  S.~Weinberg,
  Phys.\ Rev.\ Lett.\  {\bf 40} (1978) 223;
  F.~Wilczek,
  Phys.\ Rev.\ Lett.\  {\bf 40} (1978) 279.



\bibitem{Kim:1979if}
  J.~E.~Kim,
  Phys.\ Rev.\ Lett.\  {\bf 43} (1979) 103;
  M.~A.~Shifman, A.~I.~Vainshtein and V.~I.~Zakharov,
  Nucl.\ Phys.\ B {\bf 166} (1980) 493.



\bibitem{Zhitnitsky:1980tq}
  A.~R.~Zhitnitsky,
  Sov.\ J.\ Nucl.\ Phys.\  {\bf 31} (1980) 260
   [Yad.\ Fiz.\  {\bf 31} (1980) 497];
  M.~Dine, W.~Fischler and M.~Srednicki,
  Phys.\ Lett.\ B {\bf 104} (1981) 199.

\bibitem{Dias:2014osa}
  A.~G.~Dias, A.~C.~B.~Machado, C.~C.~Nishi, A.~Ringwald and P.~Vaudrevange,
  JHEP {\bf 1406} (2014) 037
  [arXiv:1403.5760 [hep-ph]].


\bibitem{Preskill:1982cy}
  J.~Preskill, M.~B.~Wise and F.~Wilczek,
  Phys.\ Lett.\ B {\bf 120} (1983) 127;
  L.~F.~Abbott and P.~Sikivie,
  Phys.\ Lett.\ B {\bf 120} (1983) 133;
  M.~Dine and W.~Fischler,
  Phys.\ Lett.\ B {\bf 120} (1983) 137;
  M.~S.~Turner and F.~Wilczek,
  Phys.\ Rev.\ Lett.\  {\bf 66} (1991) 5;
  D.~H.~Lyth and E.~D.~Stewart,
  Phys.\ Rev.\ D {\bf 46} (1992) 532.



\bibitem{Aad:2012tfa}
  G.~Aad {\it et al.}  [ATLAS Collaboration],
  Phys.\ Lett.\ B {\bf 716} (2012) 1
  [arXiv:1207.7214 [hep-ex]].
  
\bibitem{Chatrchyan:2012ufa}
  S.~Chatrchyan {\it et al.}  [CMS Collaboration],
  Phys.\ Lett.\ B {\bf 716} (2012) 30
  [arXiv:1207.7235 [hep-ex]].
  
\bibitem{Branco:2011iw}
  G.~C.~Branco, P.~M.~Ferreira, L.~Lavoura, M.~N.~Rebelo, M.~Sher and J.~P.~Silva,
  Phys.\ Rept.\  {\bf 516} (2012) 1
  [arXiv:1106.0034 [hep-ph]].



\bibitem{Glashow:1976nt}
  S.~L.~Glashow and S.~Weinberg,
  Phys.\ Rev.\ D {\bf 15} (1977) 1958;
  E.~A.~Paschos,
  Phys.\ Rev.\ D {\bf 15} (1977) 1966.
  



\bibitem{Pich:2009sp}
  A.~Pich and P.~Tuzon,
  Phys.\ Rev.\ D {\bf 80} (2009) 091702
  [arXiv:0908.1554 [hep-ph]].

  
\bibitem{Ferreira:2010xe}
  P.~M.~Ferreira, L.~Lavoura and J.~P.~Silva,
  Phys.\ Lett.\ B {\bf 688} (2010) 341
  [arXiv:1001.2561 [hep-ph]].


\bibitem{Srednicki:1985xd}
  M.~Srednicki,
  Nucl.\ Phys.\ B {\bf 260} (1985) 689;
  D.~B.~Kaplan,
  Nucl.\ Phys.\ B {\bf 260} (1985) 215.
  

\bibitem{Raffelt:2006cw}
For a review of astrophysical axion bounds see:  G.~G.~Raffelt,
  Lect.\ Notes Phys.\  {\bf 741} (2008) 51
  [hep-ph/0611350].
  
\bibitem{Bertolami:2014wua}
  M.~M.~M.~Bertolami, B.~E.~Melendez, L.~G.~Althaus and J.~Isern,
  arXiv:1406.7712 [hep-ph].
  

  
\bibitem{Minkowski:1977sc}
  P.~Minkowski,
  Phys.\ Lett.\ B {\bf 67} (1977) 421;
  M.~Gell-Mann, P.~Ramond and R.~Slansky,
  Conf.\ Proc.\ C {\bf 790927} (1979) 315
  [arXiv:1306.4669 [hep-th]];
  T.~Yanagida,
  Conf.\ Proc.\ C {\bf 7902131} (1979) 95.
  R.~N.~Mohapatra and G.~Senjanovic,
  Phys.\ Rev.\ Lett.\  {\bf 44} (1980) 912;
  J.~Schechter and J.~W.~F.~Valle,
  Phys.\ Rev.\ D {\bf 22} (1980) 2227.
  
\bibitem{Chikashige:1980ui}
  Y.~Chikashige, R.~N.~Mohapatra and R.~D.~Peccei,
  Phys.\ Lett.\ B {\bf 98} (1981) 265;
  G.~B.~Gelmini and M.~Roncadelli,
  Phys.\ Lett.\ B {\bf 99} (1981) 411;
  J.~E.~Kim,
  Phys.\ Lett.\ B {\bf 107} (1981) 69;
  P.~Langacker, R.~D.~Peccei and T.~Yanagida,
  Mod.\ Phys.\ Lett.\ A {\bf 1} (1986) 541.



  

\bibitem{Bae:2010ai}
  K.~J.~Bae,
  Phys.\ Rev.\ D {\bf 82} (2010) 055004
  [arXiv:1003.5869 [hep-ph]];
  H.~Serodio,
  Phys.\ Lett.\ B {\bf 700} (2011) 133
  [arXiv:1104.2545 [hep-ph]];
  G.~Cree and H.~E.~Logan,
  Phys.\ Rev.\ D {\bf 84} (2011) 055021
  [arXiv:1106.4039 [hep-ph]];
  I.~de Medeiros Varzielas,
  Phys.\ Lett.\ B {\bf 701} (2011) 597
  [arXiv:1104.2601 [hep-ph]].
  
\bibitem{Jung:2013hka}
  M.~Jung and A.~Pich,
  JHEP {\bf 1404} (2014) 076
  [arXiv:1308.6283 [hep-ph]];
  W.~Dekens, J.~de Vries, J.~Bsaisou, W.~Bernreuther, C.~Hanhart, U.~-G.~Mei\ss ner, A.~Nogga and A.~Wirzba,
  JHEP {\bf 1407} (2014) 069
  [arXiv:1404.6082 [hep-ph]].

\bibitem{Georgi:1981pu}
  H.~M.~Georgi, L.~J.~Hall and M.~B.~Wise,
  Nucl.\ Phys.\ B {\bf 192} (1981) 409;
  S.~Ghigna, M.~Lusignoli and M.~Roncadelli,
  Phys.\ Lett.\ B {\bf 283} (1992) 278;
  R.~Holman, S.~D.~H.~Hsu, T.~W.~Kephart, E.~W.~Kolb, R.~Watkins and L.~M.~Widrow,
  Phys.\ Lett.\ B {\bf 282} (1992) 132
  [hep-ph/9203206];
  M.~Kamionkowski and J.~March-Russell,
  Phys.\ Lett.\ B {\bf 282} (1992) 137
  [hep-th/9202003];
  S.~M.~Barr and D.~Seckel,
  Phys.\ Rev.\ D {\bf 46} (1992) 539;
  R.~Kallosh, A.~D.~Linde, D.~A.~Linde and L.~Susskind,
  Phys.\ Rev.\ D {\bf 52} (1995) 912
  [hep-th/9502069].


  
\end{thebibliography}
\end{document}